\documentstyle[twoside,fleqn,espcrc2,epsf]{article}

\newcommand{\AmS}{{\protect\the\textfont2
  A\kern-.1667em\lower.5ex\hbox{M}\kern-.125emS}}

\title{Light hadron spectrum with Kogut-Susskind quarks\thanks{Presented 
by S.~Gottlieb 
at Lattice '98, to appear.}
}

\author{C.\  Bernard,\address{Department of Physics, 
Washington University, St.\ Louis, MO 63130, USA}
        C.\ DeTar,\address{Physics Department,
University of Utah, Salt Lake City, UT 84112, USA}
        Steven Gottlieb,\address{Department of Physics,
Indiana University, Bloomington, IN 47405, USA}
        Urs\ M.~Heller,\address{SCRI, The Florida State University,
Tallahassee, FL 32306-4130, USA}
        J.E.~Hetrick,\address{University of the Pacific, 
Stockton, CA 95211-0197, USA}
	C.~McNeile,${}^{\rm b}$
        K.~Rummukainen,\address{Nordita, Blegdamsvej 17, DK-2100 
Copenhagen \O, Denmark}
        R.\ Sugar,\address{Department of Physics,
University of California, Santa Barbara, CA 93106, USA}
	and
        D.\ Toussaint,\address{Department of Physics,
University of Arizona, Tucson, AZ 85721, USA}
} 
       
\begin{document}

\begin{abstract}
We made an extensive study of the light hadron 
spectrum using the Wilson gauge action and Kogut-Susskind 
quarks.  Using both dynamical quarks and the quenched approximation,
we determine hadron masses for five and four gauge couplings
respectively, with at least   
five quark masses at each coupling.  In the continuum limit, we find
a significant difference between two flavors of dynamical quarks and
the quenched approximation for a range of quark masses.

\end{abstract}

\maketitle

Calculating the light quark spectrum is a long term goal of QCD.
Because of limitations of the numerical approach, one must work with quark
masses heavier than in Nature and with a non-zero lattice spacing.
Good control over the extrapolations in quark mass and lattice spacing
requires that simulations be carried out  
over a wide range of quark mass and lattice spacing.
We have been studying the spectrum with Kogut-Susskind or staggered quarks
for quite some time \cite{MILCTSUKUBA}.  
This year, we extended our dynamical data set
and developed a new extrapolation method for the quenched case.

In the quenched approximation, 
in the continuum limit at the physical value of $m_\pi/m_\rho$,  
we find $m_N/m_\rho = 1.254 \pm 0.018 \pm 0.028$ \cite{MILCPRL}, 
where the first error is
statistical and the second is systematic.  
With dynamical quarks, the double extrapolation in quark mass and 
lattice spacing is difficult.  However, we
can confidently extrapolate to the continuum limit at intermediate
values of the quark mass where we only need to interpolate in quark mass 
or extrapolate slightly from our lightest quark mass. 
For dynamical quarks at such intermediate quark masses,
 $m_N/m_\rho$ is significantly larger than it is in
the quenched approximation.  We believe this is the first convincing
evidence of the effects of dynamical quarks on the light quark spectrum.
Before having confidence in our extrapolation to the chiral limit, we
would need  
to have results for lighter quarks at our two weakest couplings.
At Edinburgh \cite{MILCEDINBURGH}, 
we presented our continuum extrapolation for the physical
value of $m_\pi/m_\rho$ and found it consistent with the experimental value.
However, at that time we only had four dynamical quark masses at the two
weakest couplings.  Having added a fifth mass in each case and choosing
some different fits of hadron propagators, some of the 
extrapolated masses have changed by two standard deviations.  Thus, we urge
some caution not to overreact to our dynamical results in the chiral limit.

Some details of the configuration generation and propagator calculation
may be found in Ref.~\cite{MILCTSUKUBA}.
For the dynamical runs, the minimum values of $m_\pi/m_\rho$ are 0.48 and
0.53 for $6/g^2=5.5$ and 5.6, respectively.  For the stronger couplings,
we have gone below 0.4.  

We have found the chiral extrapolation for the quenched simulations to be
a most vexing problem.  
Quenched chiral perturbation theory (Q$\chi$PT) indicates that there are 
corrections to the nucleon and $\rho$ masses that do not
occur with dynamical quarks.  
The chiral expansion is commonly expressed in terms of the quark mass.
For the quenched case, the terms $m_q^{1/2}$ and $m_q \log m_q$ are the new 
terms.  (They are due to $\eta'$ loops.)  
However, we note that due to flavor symmetry
breaking for staggered quarks, the flavor singlet pion that appears in
Q$\chi$PT does not actually have a mass proportional to $m_q^{1/2}$.  Thus,
it is actually more appropriate to express the chiral expansion using
a term proportional to a non-Goldstone pion mass.  We define, for fixed
$\lambda_N$,   
\begin{eqnarray}
m'_N \equiv (m_N + \lambda_N m_{\pi_2}){m_N^{\rm phys}
\over m_N^{\rm phys} +  \lambda_N m_{\pi}^{\rm phys}},
\label{primedmasses}
\end{eqnarray}
where phys stands for the physical values and the other quantities are
values computed at a given quark mass and lattice spacing.  A similar
equation applies to the $\rho$.  We then fit
$m'_N$ and $m'_\rho$ using only terms that appear in ordinary $\chi$PT
($M+am_q+bm_q^2$ plus either $cm_q^{3/2}$ or $c m_q^2 \log m_q$)
for various values of $\lambda_N$ and $\lambda_\rho$ obeying 
$0\le \lambda_N\le\lambda_\rho\le0.4$, which is the range expected from
Q$\chi$PT.  Further details of the chiral and continuum extrapolations
may be found in Ref.~\cite{MILCPRL}.

\begin{figure}[t]
\epsfxsize=0.96 \hsize
\epsfysize=0.96 \hsize
\epsffile{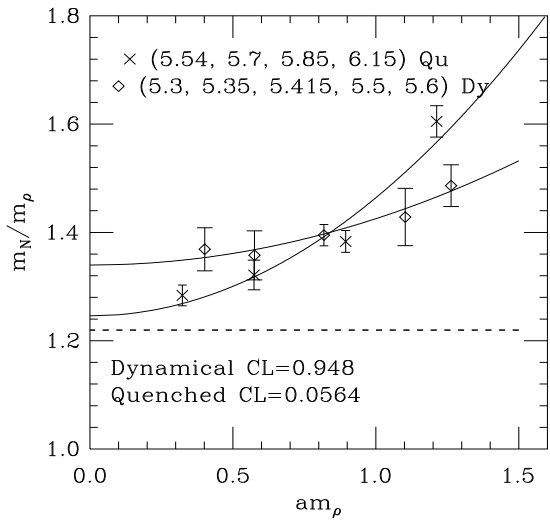}
\vspace{-38pt}
\caption{Continuum extrapolation for $m_N/m_\rho$ with the physical quark mass.}
\vspace{-17pt}
\label{fig:extrap17}
\end{figure}

\begin{figure}[thb]
\epsfxsize=0.96 \hsize
\epsfysize=0.96 \hsize
\epsffile{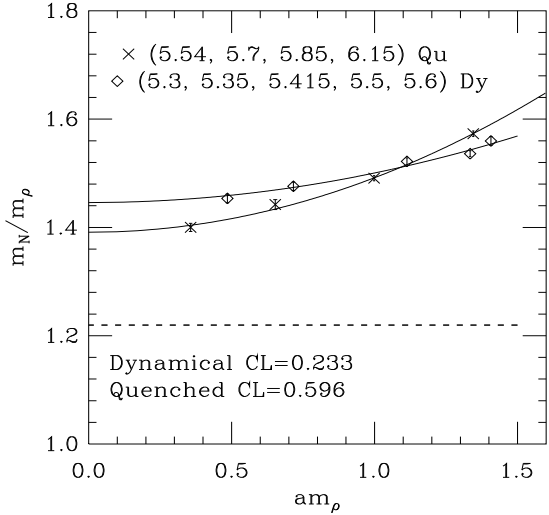}
\vspace{-38pt}
\caption{Continuum extrapolation for $m_N/m_\rho$ with $m_\pi/m_\rho=0.5$.}
\label{fig:extrap5}
\end{figure}

For the dynamical quark calculations, we tried six fits that only
contain terms appropriate to ordinary $\chi$PT.  The same four parameter forms
we used above give good combined confidence level for the nucleon
of 0.55 and 0.41.  However, for the $\rho$ the combined CL is quite poor.
The poor fits occur for the strongest couplings.  At $6/g^2=5.3$, in particular,
there are two points that lie off a smooth curve.  So far, we have picked
fits independently at each run; however, we expect that the appropriate fitting
range should be a smooth function of the quark mass.  We may adjust our fitting
ranges with this in mind.  

The final extrapolation is in the lattice spacing.
The leading error for Kogut-Suskind quarks is $O(a^2)$.  
For the quenched case, we must consider different values of $\lambda_N$
and $lambda_\rho$, and  
the systematic error comes from this variation, as well as considering
a possible quartic lattice spacing error, or a contamination from the $\pi_2$
that results in a linear error.
When we combine all these systematic errors in quadrature, we find our final
result
$m_N/m_\rho= 1.254\pm 0.018\pm0.028$.

For the dynamical quark results, 
$\lambda_N=\lambda_\rho=0$,  
so the analysis is more straightforward.  
Note, however, that the 
lattice spacing changes with quark mass.  It is therefore
helpful to fix $m_\pi/m_\rho$  (and hence the quark mass) to
various values and then to do the continuum extrapolation
separately for each choice of $m_\pi/m_\rho$.
We compare the dynamical case with the quenched by letting
$\lambda_N=\lambda_\rho=0$ 
and using the same fitting function for N and $\rho$.  (The terms specific
to Q$\chi$PT are only large for very small quark mass.)
In Figs.~1 and 2.
we compare quenched and dynamical results at $m_\pi/m_\rho=0.1753$ and 0.5
with continuum fits that include an error quadratic in the lattice spacing.

\begin{figure}[thb]
\epsfxsize=0.96 \hsize
\epsfysize=0.96 \hsize
\epsffile{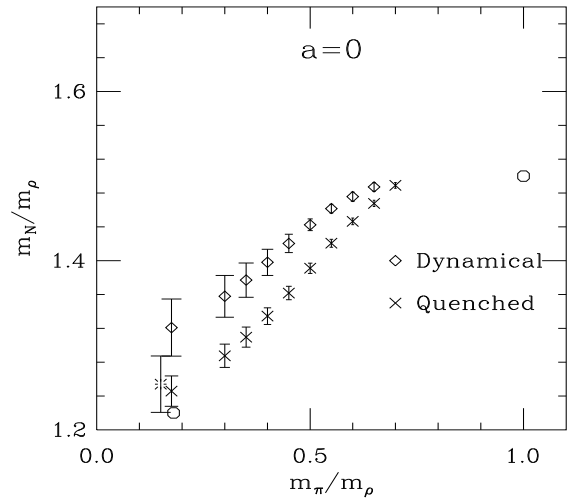}
\vspace{-38pt}
\caption{Edinburgh plot in the continuum limit for $N_f=0$ and 2.
}
\label{fig:continuumedin}
\end{figure}

In Fig.~3,
we compare the quenched and dynamical results in the continuum 
limit.  (The burst, slightly displaced to the left, includes both systematic
and statistical errors \cite{MILCPRL}.)
The dynamical results are above the quenched results for
the entire range of $m_\pi/m_\rho$ we consider.  
For $m_\pi/m_\rho=0.55$
there is clearly a significant difference between the two (0.041 $\pm$ 0.007).
This is a value where we don't have to extrapolate in quark mass.  Also,
if the quark mass were really this heavy, the $\rho$ would not decay.
We are seeing a real difference between the two flavor dynamical spectrum
and the quenched approximation.  Of course, we are especially interested
in the chiral limit, because there we can compare with the real world.
Recognizing that the quenched result for the physical value of
$m_\pi/m_\rho$ is closer to the experimental result, we are left to speculate
why this is the case.
First, we wish to again mention that for the dynamical case 
we have less confidence in
the chiral limit because of the long extrapolation for the two weaker couplings.
Second, as has always been the case with dynamical simulations, we take
no special measures to deal with the potential for $\rho$ decay.  In
lattice simulations, the threshold for decay is changed by  
finite volume
effects.  Perhaps the lattice suppression of the decay
changes the $\rho$ mass, or a complete analysis
of the $\rho$ and two $\pi$ channels is required
to correctly determine the $\rho$ mass.
Third, we only have two dynamical quarks in our calculation.  Perhaps
the strange quark plays a larger role than we might have expected.

Although much additional work remains, particularly regarding
the chiral extrapolation, we are starting to obtain reliable
results for lattice QCD, even with dynamical fermions, in the continuum 
limit.  We see a very interesting difference between the $N_f=2$ theory
and the quenched approximation even in a region where only chiral interpolation
is necessary.

This work was supported by the DOE and the NSF.  Computations were done at
CTC, Indiana University, NCSA, ORNL, PSC, Sandia, SCRI and SDSC.

\end{document}